\documentclass[12pt,letterpaper]{article}

\usepackage[utf8]{inputenc}
\usepackage[T1]{fontenc}
\usepackage[margin=1in]{geometry}
\usepackage{amsmath,amssymb,amsfonts}
\usepackage{graphicx}
\usepackage{float}
\usepackage{placeins}
\usepackage{needspace}
\usepackage{booktabs}
\usepackage[font=small,skip=6pt]{caption}
\usepackage{authblk}
\usepackage{parskip}
\usepackage{natbib}
\usepackage{xcolor}
\usepackage{hyperref}

\definecolor{linkcolor}{RGB}{0,90,160}

\hypersetup{
  colorlinks=true,
  linkcolor=linkcolor,
  urlcolor=linkcolor,
  citecolor=linkcolor
}

\title{Opponent-Adjusted Evaluation of NFL Pass Protection and Pass Rush}
\author{Jonathan Pipping-Gam\'on\thanks{These authors share first authorship.\newline\hspace*{1.8em}Corresponding author: Jonathan Pipping-Gam\'on (\href{mailto:jpipping@wharton.upenn.edu}{jpipping@wharton.upenn.edu}).},
  Maximilian Gebauer\protect\footnotemark[1],
  Victoria Lee\protect\footnotemark[1],
  Kenny Watts, Abraham J. Wyner}
\affil{University of Pennsylvania}
\date{}

\begin{document}
\maketitle


\begin{abstract}
Evaluating NFL pass protectors and rushers is difficult because observed win rates reflect both player performance and the difficulty of the blocking assignments each player faces.
Using Hudl player coordinates and blocking annotations from the 2021 NFL regular season, we estimate blocker and rusher strength jointly from initial blocking matchups.
A ridge-regularized Bradley--Terry model uses early pass rush wins and losses within 2.5 seconds of the snap, while a multinomial extension models full-play losses, wins, pressures, and sacks.
To summarize full-play performance, we use the outcomes' conditional associations with offensive expected points added (EPA) to weight predicted probabilities, giving ratings of expected disruption produced by rushers or allowed by blockers.
In held-out Weeks 16--18, the early and final models reduce log loss by 4.5\% and 3.6\%, respectively, relative to historical matchup baselines whose smoothing strengths are selected by cross-validation; 95\% game-bootstrap intervals for both reductions are above zero.
The resulting ratings place observed success in the context of opponents and protection help, with game-bootstrap intervals quantifying sampling uncertainty in early strength and full-play disruption.
\end{abstract}

\section{Introduction}\label{sec:introduction}
Evaluating offensive line performance in pass protection remains an open problem in football analytics.
The central challenge is to distinguish the quality of protection from the actions of the quarterback, his receivers, and the opposing defense, all of which influence the result of a passing play.
A quarterback may complete a quick pass despite a missed block, while a sack can occur after his blockers have sustained protection for several seconds.
The result of the play therefore cannot, on its own, tell us how well the line performed, making it difficult to assess the unit or compare the performance of individual blockers \citep{nguyen2023}.

Player tracking data allow us to examine the blocking contests within a play, bringing the evaluation closer to the work of individual linemen.
Pass block win rate, for example, records whether a blocker sustains protection over a specified time window \citep{burke2018}.
This rate provides a useful description of observed success, but its interpretation depends on each player's underlying assignments: a lineman facing strong defenders without help faces a more difficult task than one who faces weaker opponents or shares a block with a teammate.
Comparing linemen therefore requires accounting for opponent strength and protection help.

Establishing a common standard for those comparisons requires evaluating opposing pass rushers as well.
Their own success depends on whom they face, so using their raw win rates to measure opponent strength leaves the original problem unresolved.
Over a season, however, these matchups link players through shared opponents: a blocker's results inform the strengths of the rushers he encounters, while their contests elsewhere help us estimate his strength.
Combining evidence across these matchups allows us to estimate strength on both sides jointly, with players compared through their expected performance against common opposition and protection help.

\subsection{Evaluating Pass Protection and Pass Rush}\label{sec:intro_existing}
Early research on offensive linemen used manually charted plays to connect pass protection with passing success.
For example, \citet{alamar2008} recorded whether each lineman sustained his block and how much time the quarterback had to throw, then examined how successful blocking was associated with pass completion.

Player tracking data later provided another way to evaluate individual blocks by recording how blockers and defenders move throughout a play.
Using these data, ESPN developed pass block and pass rush win rates: a blocker wins by sustaining protection for 2.5 seconds, and a rusher wins by beating a block within that interval \citep{burke2018}.
The resulting rates describe how often players succeed in their blocking matchups, but they also reflect the difficulty of the assignments each player faced.
Comparing players therefore requires accounting for opponent strength and help from other blockers.

While win rates summarize whether a block was held or beaten, tracking data can also show how close a defender came to the quarterback during the play.
\citet{nguyen2023} use this information to construct STRAIN, a measure based on the defender's distance from the quarterback and how quickly that distance is shrinking.
The measure therefore captures the progress of a rush even when it produces no recorded sack or hurry.
To compare players, the authors then model play-level average STRAIN while accounting for the opposing blocker, the offensive and defensive teams, player positions, and the situation on the play.
When more than one blocker is involved, they use the nearest blocker at the start of the play.

\subsection{Our Approach and Contributions}\label{sec:intro_contribution}
We estimate blocker and rusher strength jointly from the results of pass blocking matchups.
Using player coordinates and engagement records supplied by Hudl from the 2021 NFL regular season, we identify each rusher's initial blocking matchup and the blockers who share that assignment.
We record an early pass rush win when the rusher moves closer to the quarterback than every blocker sharing the initial assignment within 2.5 seconds of the snap and before pass release or a sack.
This uses the same time threshold as pass block and pass rush win rates \citep{burke2018}; when the matchup is sufficiently observed and no such win occurs, we record an early pass rush loss.
We fit a paired-comparison model by penalized maximum likelihood, accounting for protection help.

Beating a block, however, does not necessarily lead to pressure on the quarterback, and players who win equally often may produce different forms of disruption.
We therefore extend the framework to four final pass rush outcomes over the full play: loss, win, pressure, and sack.
To summarize differences across these outcomes, we estimate how each is associated with offensive expected points added (EPA), accounting for the recorded situation on the play.
We use those estimated values to weight the predicted outcome probabilities, giving a measure of expected disruption produced by a rusher or allowed by a blocker.

We evaluate both models by predicting Weeks 16--18 from data in Weeks 1--15 and comparing their predictions with baselines based on players' historical outcome rates.
Full-season comparisons also show how the model's rankings differ from those based on observed early pass rush win rates, and how including pressures and sacks changes the evaluation.
Game-level bootstraps quantify uncertainty in both prediction and player comparisons.

Section~\ref{sec:data} describes the data and outcomes, Section~\ref{sec:models} develops the models and ratings, and Section~\ref{sec:validation} explains estimation and validation.
We present the results in Section~\ref{sec:results} and discuss their implications and limitations in Section~\ref{sec:discussion}.

\section{Data and Outcome Definitions}\label{sec:data}
\subsection{Blocking Assignments}\label{sec:protection}
We define each observation as a rusher's initial blocking matchup, using the start and end times in Hudl's blocking annotations to identify the first engagement.
The first blocker may be an offensive lineman, tight end, running back, or wide receiver, and we retain that assignment regardless of position.
If another blocker's recorded block overlaps the ongoing matchup for a positive interval before the rusher wins, he shares the assignment and is included in the same observation.

The overlap requirement identifies help while the initial block is still underway.
Once the initial matchup ends, any later pickup or new block after a gap belongs to a separate assignment.

A play from Pittsburgh's 2021 season opener at Buffalo illustrates how we record this distinction.
Cameron Heyward sacked Josh Allen on third-and-six in the fourth quarter \citep{pitbuf2021}, and the Hudl blocking record shows him first engaging Mitch Morse, followed by Jon Feliciano.
Because Feliciano's recorded block begins exactly when Morse's ends, their intervals have no period of overlap.
We therefore record Heyward's initial assignment as a matchup against Morse alone, without protection help; Feliciano's block is a later pickup.

\subsection{Early and Final Pass Rush Outcomes}\label{sec:outcomes}
To grade initial matchups, we examine Hudl's recorded event snapshots within 2.5 seconds of the snap, while the matchup is ongoing and strictly before pass release or a sack.
We record an early pass rush win when, in the same snapshot, the rusher is closer to the quarterback than every blocker who has joined his assignment.
Let $d_{a,t}$ denote the straight-line distance from player $a$ to the quarterback in snapshot $t$; for rusher $r$, we require
\[
d_{r,t}<d_{b,t}\qquad\text{for every blocker }b\text{ who has joined the assignment by }t.
\]

When the relevant observations are complete and contain no qualifying win, we record an early pass rush loss.
On a quick pass, for example, an early pass rush loss means that no qualifying win was observed before pass release.
If missing positions, uncertain event timing, or unclear assignment boundaries prevent us from determining whether the rusher won, we leave the matchup ungraded.

We also follow each rusher through the play to record pressures from Hudl's full-play annotations and sacks from official credits.
We assign four final categories: sack when the rusher receives sack credit, pressure when he records pressure without sack credit, and otherwise win or loss according to his early pass rush grade.
A credited half sack receives the sack category, with its credit fraction retained separately.

To compare all four final categories on the same basis, we require an observed early pass rush grade, verified pressure coverage for the game, and reconciled sack credits for the play.
An observed early pass rush grade remains usable even when those records are incomplete.

On the Buffalo play, Heyward has no observed early pass rush win against Morse but receives sack credit, while T.J.\ Watt wins his initial matchup against Daryl Williams and finishes in the win category (Table~\ref{tab:buffalo_outcomes}).

\begin{table}[htbp]
\centering
\caption{Rusher outcomes on the Buffalo play described in Section~\ref{sec:protection}. The early pass rush result concerns the initial blocking matchup, while the full-play category follows the rusher through the play.}
\label{tab:buffalo_outcomes}
\begin{tabular}{llll}
\toprule
Rusher & Initial blocker & Early pass rush result & Full-play category \\
\midrule
Cameron Heyward & Mitch Morse & Loss & Sack \\
T.J.\ Watt & Daryl Williams & Win & Win \\
Tyson Alualu & Cody Ford & Win & Pressure \\
Melvin Ingram & Dion Dawkins & Loss & Loss \\
\bottomrule
\end{tabular}
\end{table}

\subsection{Sample Construction}\label{sec:data_sample}
The Hudl data span Weeks 1--18 of the 2021 regular season, covering 266 games and 20,449 plays labeled as passes or sacks.
The 2023 NFL Big Data Bowl similarly pairs player coordinates with scouting annotations, but covers only Weeks 1--8 of the 2021 season \citep{bdb2023}.
The temporal resolution is also comparable within our initial blocking windows: the Big Data Bowl's Next Gen Stats positions are recorded every 0.1 seconds \citep{nguyen2023}, and Hudl's event-associated snapshots are generally no farther apart.
Hudl additionally supplies block start and end annotations for the blocking assignments.
We link these records to nflverse play-by-play to obtain EPA and play context, and to verify which defenders received full or shared sack credit \citep{nflverse2026}.
A link is accepted when the game information, clock, play sequence, and recorded players and events identify a unique match; 19,583 plays (95.8\%) meet this rule.

We construct each analysis sample from the linked plays, retaining initial blocking matchups for defensive linemen and linebackers whose blockers can be identified and whose early pass rush outcome can be graded.
We exclude nullified or deleted plays, two-point conversions, and records with inconsistent player information.
The early pass rush model uses 70,304 matchups from 18,225 plays.

For the final pass rush model, we exclude four games without verified pressure coverage and matchups with unresolved sack credits.
Table~\ref{tab:sample_accounting} summarizes the resulting samples.

\begin{table}[htbp]
\centering
\caption{Analysis samples after linkage and eligibility checks. Each matchup is one rusher's initial blocking assignment.}
\label{tab:sample_accounting}
\begin{tabular}{lrrr}
\toprule
Sample & Games & Plays & Matchups \\
\midrule
Early pass rush outcome & 266 & 18,225 & 70,304 \\
Final pass rush outcome & 262 & 17,920 & 69,114 \\
\bottomrule
\end{tabular}
\end{table}


\section{Opponent-Adjusted Models and Player Ratings}\label{sec:models}

\subsection{Early Pass Rush Model}\label{sec:model_early}
When a rusher faces a single blocker, we model the probability of an early pass rush win through the difference between their strengths.
This is the Bradley--Terry paired-comparison model \citep{bradley1952}, which we extend to shared blocking assignments by combining the participating blockers' strengths and adding an adjustment for protection help.

Let $Y_k$ indicate an early pass rush win in matchup $k$, with probability $p_k=\Pr(Y_k=1)$.
Denote the rusher by $i(k)$, the blockers in his initial assignment by $J_k$, and their number by $m_k=|J_k|$.
The indicator $D_k$ records overlapping help, as defined in Section~\ref{sec:protection}.
We model
\begin{equation}\label{eq:early_model}
\operatorname{logit}(p_k)
=\alpha+r_{i(k)}-\frac{1}{m_k}\sum_{j\in J_k}b_j+\delta D_k.
\end{equation}
Larger $r_i$ indicates greater rusher strength and increases the win probability, while larger $b_j$ indicates better protection and reduces it.

The mean blocker effect gives each blocker equal weight and keeps protection strength on the same scale for single and shared assignments.
The term $\delta D_k$ then represents the association with overlapping help after accounting for the players involved.
Each matchup contributes one binary observation, regardless of the number of participating blockers.

We estimate all player and help effects jointly by ridge-penalized maximum likelihood, using the outcomes across matchups to learn the strengths on both sides.
The penalty stabilizes estimates for players with limited observations; Section~\ref{sec:validation} describes its selection.
For reporting, we center player effects at zero within each role and adjust $\alpha$ to preserve fitted probabilities.
The intercept then gives the baseline log odds at zero player effects and no recorded help.

\subsection{Final Pass Rush Model}\label{sec:model_final}
Rushers with similar chances of an early pass rush win may differ in how often their rushes end in pressure or a sack.
We therefore extend the same blocking-assignment structure to a multinomial model, allowing the player and help effects to differ across final pass rush outcomes.

Let $C_k$ denote the rusher's final pass rush outcome in matchup $k$, taking values in $\mathcal{C}=\{\mathrm{loss},\mathrm{win},\mathrm{pressure},\mathrm{sack}\}$ as defined in Section~\ref{sec:outcomes}.
Using the notation from the early pass rush model, we specify
\begin{equation}\label{eq:final_model}
\begin{aligned}
\eta_{k,c} &= \alpha_c+r_{i(k),c}-\frac{1}{m_k}\sum_{j\in J_k}b_{j,c}+\delta_cD_k, \\
\Pr(C_k=c) &= \frac{\exp(\eta_{k,c})}{\sum_{d\in\mathcal{C}}\exp(\eta_{k,d})},\qquad c\in\mathcal{C}.
\end{aligned}
\end{equation}

The player coefficients form profiles across outcomes.
For a rusher, a larger difference $r_{i,c}-r_{i,d}$ increases the odds of category $c$ relative to category $d$ for the same blocking assignment.
Blocker effects enter with the opposite sign, so a larger corresponding difference $b_{j,c}-b_{j,d}$ reduces those odds.

As in the early pass rush model, we estimate all player and help effects jointly with ridge penalization.
For reporting, each coefficient profile sums to zero across categories, and player effects remain centered within roles, with intercept adjustments that preserve predicted probabilities.
We next use offensive EPA to compare the value of these outcomes.

\subsection{Valuing Disruption}\label{sec:valuation}
Beating a block, creating pressure, and recording a sack have different values to the defense.
To weight these outcomes appropriately in player rankings, we estimate their value using offensive EPA, which measures how a play changes the offense's expected points.

Several defenders can record rushing outcomes on the same play, but those outcomes share a single EPA value.
We therefore enter each play once in a linear regression, with its rushing outcomes included together.
We fit the regression on 13,231 linked plays from Weeks 1--15 with observed EPA, play context, and complete early and final pass rush outcomes for every observed protection participant (Section~\ref{sec:outcomes}).

For play $\ell$, let $W_\ell$ and $P_\ell$ count rushers whose final pass rush outcome is win or pressure, respectively, and let $S_\ell$ denote total sack credit.
The sack total uses credited fractions, so two half sacks contribute one full sack.
We model
\begin{equation}\label{eq:epa_model}
\mathrm{EPA}_\ell
=\beta_0+\beta_{\mathrm{win}}W_\ell+\beta_{\mathrm{pressure}}P_\ell+\beta_{\mathrm{sack}}S_\ell
+f(\mathbf{z}_\ell)+\varepsilon_\ell.
\end{equation}
The term $f(\mathbf{z}_\ell)$ collects linear adjustments for play context, quarterback, defensive team, the number of observed protection participants, and shared sacks.
The full regression specification and outcome accounting are given in Appendix~\ref{app:epa_model}.

Let $\widehat{\beta}_c$ denote the fitted EPA coefficient for outcome $c$, with $\widehat{\beta}_{\mathrm{loss}}=0$.
Since lower offensive EPA favors the defense, the corresponding disruption value is $-\widehat{\beta}_c$.
We express these values relative to a sack as
\begin{equation}\label{eq:disruption_weights}
w_c=\frac{\widehat{\beta}_c}{\widehat{\beta}_{\mathrm{sack}}},\qquad
w_{\mathrm{loss}}=0,\qquad w_{\mathrm{sack}}=1.
\end{equation}
These weights reflect EPA associations after accounting for other outcomes and play context.

\subsection{Player Ratings}\label{sec:ratings}
For early pass rush wins and losses, we rank rushers and blockers by their fitted strength coefficients, $\widehat{r}_i$ and $\widehat{b}_j$, respectively.
Within either role, more positive coefficients indicate greater strength, with zero corresponding to the mean fitted player effect.

We summarize full-play disruption by estimating how much each player would produce or allow under the season's observed mix of blocking assignments.
After fitting, we evaluate every player over the same fixed reference within his role and combine the predicted outcome probabilities with their EPA-based disruption weights, $w_c$, from Section~\ref{sec:valuation}.
Each rusher faces the same mix of blocking groups and help.
For a blocker, we predict outcomes with him in place of a member of each blocking group, keeping the opposing rusher, the other blockers, and help fixed.
When a group has multiple members, we make this prediction for each member in turn and average the results.
Each observed matchup then receives equal weight in the final average.

Let $\bar{p}_{a,c}$ denote the resulting average probability of outcome $c$ for player $a$.
The final-outcome rating is
\begin{equation}\label{eq:player_rating}
Q_a=\sum_{c\in\mathcal{C}}w_c\bar{p}_{a,c}.
\end{equation}
We therefore rank rushers from highest to lowest $Q_a$ and blockers from lowest to highest.

\section{Estimation and Validation}\label{sec:validation}
\subsection{Fitting and Tuning}\label{sec:estimation_tuning}
We fit both pass rush models in R using the \texttt{glmnet} package \citep{friedman2010}.
We choose a ridge penalty for each pass rush model by five-fold cross-validation, keeping all matchups from a game in the same fold.
For each candidate penalty, we fit on four folds and predict outcomes in the fifth, repeating the process across the five folds.
The penalty giving the lowest average cross-validation log loss is then used to refit the model on all five folds.

For predictive evaluation, we restrict this procedure to Weeks 1--15 and use the fitted models to predict Weeks 16--18 (Section~\ref{sec:predictive_evaluation}).
The season ratings draw on all eligible games from Weeks 1--18, so we repeat tuning and fitting on that sample while retaining the EPA weights estimated from the complete Weeks 1--15 plays (Section~\ref{sec:valuation}).
Appendix~\ref{app:cv_tuning} shows the cross-validation curves for both fitting samples.

\subsection{Predictive Evaluation}\label{sec:predictive_evaluation}
Using the training-period fits from Section~\ref{sec:estimation_tuning}, we predict outcomes for matchups in Weeks 16--18 and compare them with two baselines built from Weeks 1--15 outcomes.
The league-average baseline uses the overall training frequencies, assigning the same outcome probabilities to every matchup.

The second baseline adds the players' histories, using the rusher's and initial blockers' outcome frequencies smoothed toward the league rates.
We select the smoothing strength by five-fold game-grouped cross-validation within Weeks 1--15, then combine the player profiles into matchup probabilities, including all blockers who share the assignment.
The construction and candidate strengths are given in Appendix~\ref{app:predictive_baselines}.

Within each model, we evaluate all three methods on the same eligible test matchups using average log loss.
We report the reduction in log loss relative to each baseline as a percentage of that baseline's loss.

\subsection{Uncertainty}\label{sec:uncertainty}
We use 1,000 game-bootstrap draws for predictive comparisons and another 1,000 for season ratings, sampling games with replacement and keeping all of their observations together.
Training and test games are sampled separately for predictive comparisons, while the season rating draws sample from the original 266 games.

In each draw, we retune and refit both pass rush models and re-estimate the EPA weights using the sampled Weeks 1--15 games.
For the primary predictive comparison, we rebuild the baseline histories within each draw while retaining the smoothing strengths selected from the original training games.
We also repeat the comparison with smoothing reselected within each draw's training games.
The player pool and common assignment reference remain fixed for the rating draws.

We report 95\% percentile intervals for the loss reductions, player scores, and ranks, recomputing ranks within each role in every rating draw.
These intervals describe variation in the retuned, regularized estimates across resampled games.

\subsection{Player Comparisons}\label{sec:player_comparisons}
We examine how the full-season model rankings differ from rankings based on observed outcome rates and their smoothed versions.
For initial blocking matchups, we use each rusher's early pass rush win rate or each blocker's corresponding protection rate; to compare full-play disruption, we combine final pass rush outcome frequencies with the same EPA weights used in Equation~\ref{eq:player_rating}.
For these descriptive season comparisons, we smooth each player's full-season history using the strengths selected by training cross-validation: 2 for early outcomes and 10 for final outcomes (Appendix~\ref{app:predictive_baselines}).

Our primary comparisons include defensive linemen and linebackers as rushers and offensive linemen as blockers.
Within each role, we rank the same players across methods, requiring at least one eligible matchup and a score from every method.
We repeat the comparisons with minimum exposures of 100, 200, and 400 matchup appearances, summarizing agreement through Spearman correlation and mean absolute rank changes, alongside individual changes.
Comparing raw and smoothed rankings helps distinguish the effect of smoothing from the further changes introduced by the model.

We also compare the rankings with the 2021 Associated Press All-Pro selections \citep{apallpro2021}, using first-team selections and, separately, selections to either team.
Within the same player cohorts, we summarize alignment by AUC and the number of selected players among the top ten.
Because these honors recognize broader football responsibilities, we treat their alignment with the rankings as a descriptive check.

Finally, we examine how the rankings change under a proportional-odds model that orders the final pass rush outcomes as loss, win, pressure, and sack.
This model uses one strength coefficient per player across the outcome thresholds, in place of the multinomial model's category-specific effects.
We tune and fit it on the full season and calculate ratings using the same EPA weights and reference assignments.
Appendix~\ref{app:ordinal_model} gives its specification and comparison populations.

\subsection{Weekly Rating Paths}\label{sec:weekly_paths}
We trace how player ratings develop as games accumulate by refitting both pass rush models on eligible matchups through each week.
The paths are retrospective: weekly fits retain the season fit's ridge penalties, player pool, assignment references, and EPA weights.

For early pass rush wins and losses, we plot the centered strength coefficients $\widehat{r}_i$ and $\widehat{b}_j$ from Section~\ref{sec:ratings}.
The full-play paths show expected disruption produced by rushers or allowed by blockers, using $Q_a$ from Equation~\ref{eq:player_rating}.

We also construct weekly paths within each season bootstrap draw from Section~\ref{sec:uncertainty}: sampled games enter at their original weeks, while the draw's selected penalties and EPA weights remain fixed.
The resulting paths give 95\% pointwise percentile intervals at each week where all 1,000 bootstrap estimates are defined.

\section{Results}\label{sec:results}
\subsection{Predictive Performance}\label{sec:results_prediction}
Both pass rush models improve prediction over the league-average and historical matchup baselines in Weeks 16--18 (Table~\ref{tab:predictive_performance}).
Against the league-average baseline, the early pass rush model reduces log loss by 8.8\% and the final pass rush outcome model by 6.1\%.
With smoothing fixed at the original training-CV choices of 2 and 10, the historical baseline provides a stronger comparison, yet the fitted models reduce its log loss by 4.5\% and 3.6\%, respectively.
The 95\% game-bootstrap intervals remain above zero, both at these fixed strengths and when smoothing is reselected within each bootstrap training sample.

\begin{table}[htbp]
\centering
\small
\caption{Held-out log loss over 47 games in Weeks 16--18, with percentage reductions and 95\% game-bootstrap percentile intervals. The primary fixed historical baselines retain the original training-CV choices of $\tau=2$ for early outcomes and $\tau=10$ for final outcomes, while retuned historical baselines reselect smoothing within each bootstrap draw.}
\label{tab:predictive_performance}
\begin{tabular}{lrrrc}
\toprule
Baseline & Model loss & Baseline loss & Reduction (\%) & 95\% interval (\%) \\
\midrule
\multicolumn{5}{l}{\textit{Early pass rush wins and losses ($n=12{,}205$ matchups)}} \\
League average & 0.448 & 0.492 & 8.8 & $[7.0,\;9.0]$ \\
Fixed ($\tau=2$) & 0.448 & 0.469 & 4.5 & $[3.4,\;5.2]$ \\
Retuned historical & 0.448 & 0.469 & 4.5 & $[3.4,\;5.0]$ \\
\midrule
\multicolumn{5}{l}{\textit{Final pass rush outcomes ($n=12{,}198$ matchups)}} \\
League average & 0.641 & 0.682 & 6.1 & $[4.7,\;6.3]$ \\
Fixed ($\tau=10$) & 0.641 & 0.665 & 3.6 & $[3.2,\;4.4]$ \\
Retuned historical & 0.641 & 0.665 & 3.6 & $[3.1,\;4.2]$ \\
\bottomrule
\end{tabular}
\end{table}

\FloatBarrier
\subsection{Early Pass Rush Ratings}\label{sec:results_ratings}

Full-season early pass rush ratings place several of the league's recognized leading performers near the top (Table~\ref{tab:early_player_ratings}).
Aaron Donald and Robert Quinn lead the rusher ratings, while Kevin Zeitler and Gabe Jackson have the highest ratings among offensive linemen.
All five rushers and three of the five linemen shown in the table received 2021 All-Pro recognition \citep{apallpro2021}.

\begin{table}[H]
\centering
\small
\caption{The five highest-rated rushers and offensive linemen for full-season early pass rush wins and losses, with 95\% game-bootstrap percentile intervals for strength ratings and ranks. Players have at least 200 eligible matchups.}
\label{tab:early_player_ratings}
\begin{tabular}{lrrcrc}
\toprule
Player & Matchups & Rating & 95\% interval & Rank & 95\% rank interval \\
\midrule
\multicolumn{6}{l}{\textit{Rushers ($n=139$)}} \\
Aaron Donald & 489 & $0.903$ & $[0.534,\;1.002]$ & 1 & $[1.0,\;9.0]$ \\
Robert Quinn & 298 & $0.848$ & $[0.496,\;1.014]$ & 2 & $[1.0,\;10.0]$ \\
Micah Parsons & 210 & $0.802$ & $[0.475,\;0.888]$ & 3 & $[1.0,\;14.0]$ \\
T.J.\ Watt & 315 & $0.721$ & $[0.437,\;0.878]$ & 4 & $[1.0,\;16.0]$ \\
Myles Garrett & 418 & $0.696$ & $[0.423,\;0.853]$ & 5 & $[1.0,\;17.0]$ \\
\midrule
\multicolumn{6}{l}{\textit{Offensive linemen ($n=177$)}} \\
Kevin Zeitler & 581 & $1.192$ & $[0.683,\;1.268]$ & 1 & $[1.0,\;20.0]$ \\
Gabe Jackson & 431 & $1.120$ & $[0.581,\;1.145]$ & 2 & $[1.0,\;31.0]$ \\
Zack Martin & 489 & $1.041$ & $[0.573,\;1.067]$ & 3 & $[1.0,\;29.0]$ \\
Lane Johnson & 285 & $1.036$ & $[0.506,\;1.173]$ & 4 & $[1.0,\;45.0]$ \\
Trent Williams & 380 & $1.035$ & $[0.570,\;1.095]$ & 5 & $[1.0,\;35.0]$ \\
\bottomrule
\end{tabular}
\par\smallskip
\begin{minipage}{\linewidth}
\footnotesize
Ratings are centered over all fitted players within each role.
Ranks and their intervals use the fixed player cohorts shown above.
\end{minipage}
\end{table}

\begin{figure}[H]
\centering
\includegraphics[width=\linewidth]{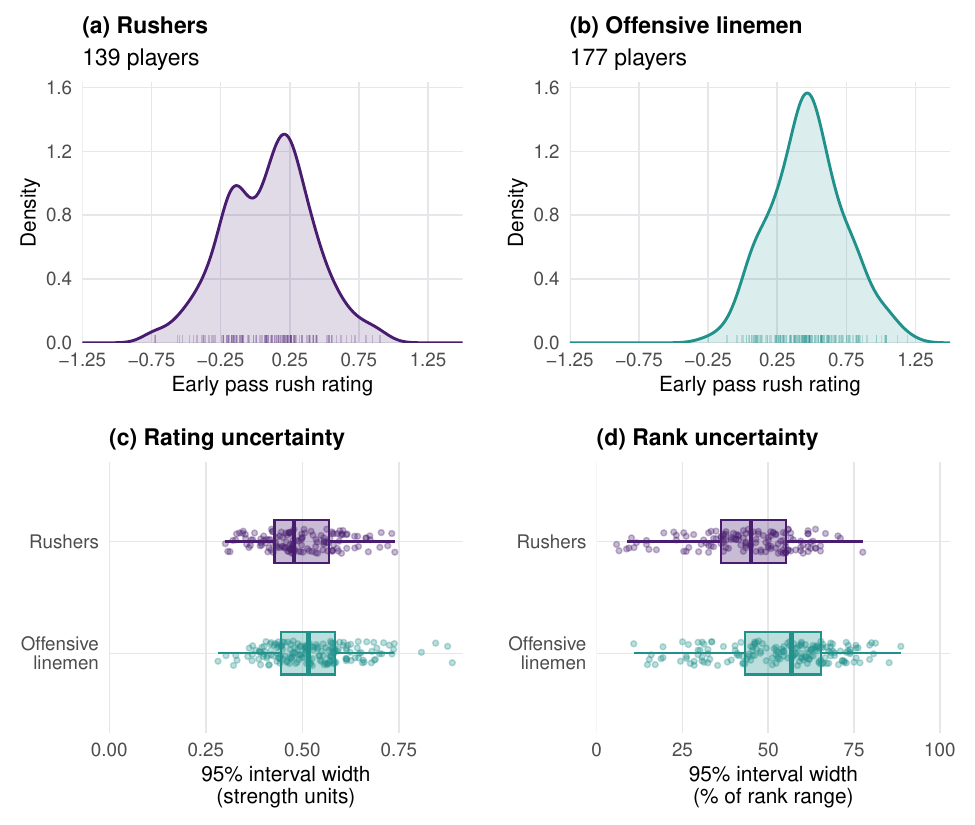}
\caption{Full-season early pass rush ratings for players with at least 200 eligible matchups. Rating distributions for (a) rushers and (b) offensive linemen, followed by distributions of players' 95\% interval widths for (c) strength ratings and (d) ranks. Rank widths are expressed as percentages of each cohort's available rank range. Ratings are centered over all fitted players within each role.}
\label{fig:early_rating_comparison}
\end{figure}

The rankings are more stable for rushers than for offensive linemen, even though rushers generally have fewer observed matchups (Figure~\ref{fig:early_rating_comparison}).
Individual ratings have similar uncertainty in the two roles, but the linemen's estimates cluster closer together through the middle of the distribution, helping to explain the greater uncertainty in their relative order.
Among the leading players, Donald's 95\% rank interval spans first through ninth, whereas Zeitler's extends from first through twentieth.

\begin{figure}[H]
\centering
\includegraphics[width=0.90\linewidth]{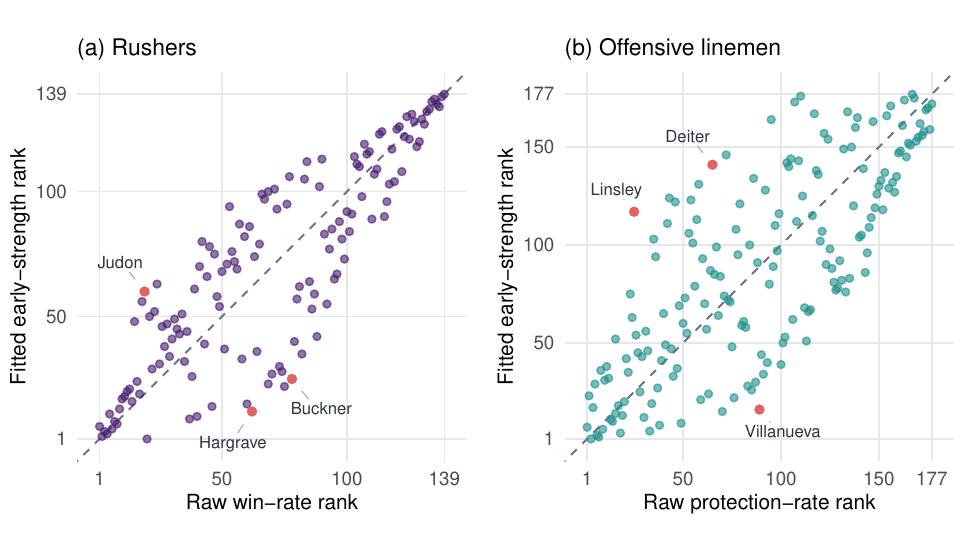}
\caption{Raw early outcome ranks versus fitted early-strength ranks for (a) rushers and (b) offensive linemen with at least 200 eligible matchups. Each panel uses the same players for both rankings, with rank one best and average ranks for ties. Spearman rank correlations are $0.85$ for rushers and $0.77$ for offensive linemen. Raw rusher rates count each matchup once; raw protection rates divide each shared assignment's count equally among its blockers, following Appendix~\ref{app:predictive_baselines}. The dashed line marks unchanged rank; highlighted players illustrate substantial rank changes.}
\label{fig:raw_adjusted_early_ranks}
\end{figure}

Our framework preserves much of the ordering suggested by observed win and protection rates, while giving substantially different assessments of some players.
The raw and fitted early rankings have Spearman correlations of 0.85 for rushers and 0.77 for offensive linemen (Figure~\ref{fig:raw_adjusted_early_ranks}).
For example, DeForest Buckner rises from 78th by raw win rate to 25th by fitted strength, while Corey Linsley moves from 25th by raw protection rate to 117th.

\subsection{Full-Play Disruption Ratings}\label{sec:results_disruption}
The EPA regression estimates the disruption value of a pressure at about one-tenth that of a sack (Table~\ref{tab:disruption_weights}).
An early pass rush win that produces neither outcome receives a much smaller weight, with a 95\% interval that includes zero.

\begin{table}[H]
\centering
\small
\caption{EPA-based disruption weights relative to a sack, with 95\% game-bootstrap percentile intervals. Loss and sack have fixed weights of zero and one, respectively.}
\label{tab:disruption_weights}
\begin{tabular}{lrc}
\toprule
Final pass rush outcome & Weight & 95\% interval \\
\midrule
Loss & $0$ & --- \\
Win only & $0.009$ & $[-0.009,\;0.025]$ \\
Pressure & $0.103$ & $[0.071,\;0.140]$ \\
Sack & $1$ & --- \\
\bottomrule
\end{tabular}
\end{table}

\begin{figure}[H]
\centering
\includegraphics[width=0.88\linewidth]{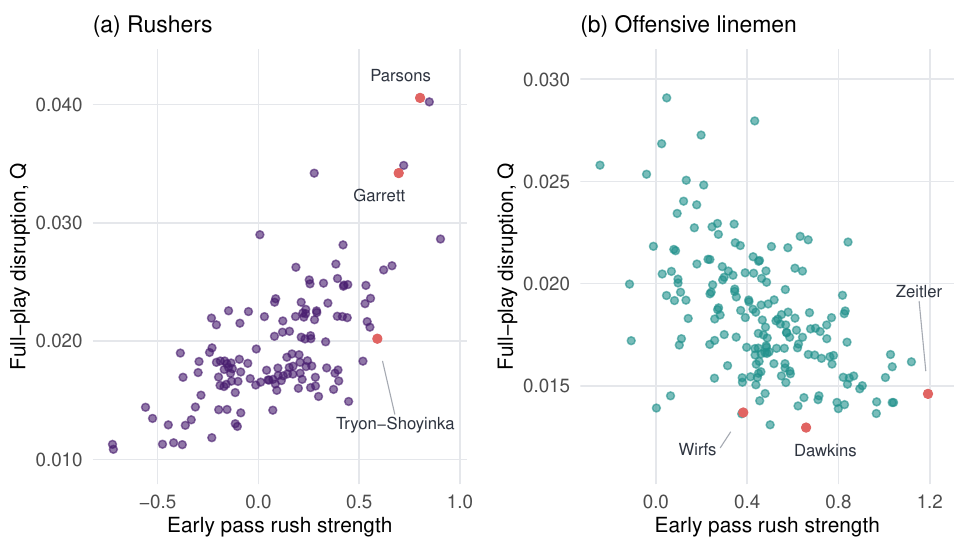}
\caption{Early pass rush strength and expected full-play disruption, $Q$, for (a) rushers and (b) offensive linemen with at least 200 eligible matchups in each model. Both axes show full-season point estimates for the same players within each role. Higher early strength is favorable in both roles; higher $Q$ favors rushers and lower $Q$ favors blockers. The Spearman correlation between early strength and $Q$ is $0.61$ for rushers and $-0.53$ for offensive linemen. Selected players are labeled to illustrate differences between the two ratings.}
\label{fig:early_disruption_comparison}
\end{figure}

Evaluating early blocking success and valuing full-play disruption give related but distinct assessments of players.
Four of the five leading rushers for initial matchup strength also rank among the five leaders for full-play disruption, whereas the two models share no top-five offensive linemen (Table~\ref{tab:disruption_player_ratings}).
Across players with at least 200 eligible matchups in both models, the Spearman correlations between the rankings for early pass rush strength and full-play disruption are 0.61 for rushers and 0.53 for offensive linemen (Figure~\ref{fig:early_disruption_comparison}).

\begin{table}[H]
\centering
\small
\renewcommand{\arraystretch}{0.95}
\caption{The five leading rushers and offensive linemen for full-play disruption, with 95\% game-bootstrap percentile intervals for $Q_a$ and ranks for initial matchup strength. Both rankings use the same players with at least 200 eligible matchups in each model. Higher $Q_a$ favors rushers; lower $Q_a$ favors blockers.}
\label{tab:disruption_player_ratings}
\begin{tabular}{lrrcr}
\toprule
Player & \shortstack{Disruption\\rank} & $Q_a$ & 95\% interval & Early rank \\
\midrule
\multicolumn{5}{l}{\textit{Rushers ($n=138$)}} \\
Micah Parsons & 1 & $0.0405$ & $[0.0241,\;0.0442]$ & 3 \\
Robert Quinn & 2 & $0.0402$ & $[0.0229,\;0.0442]$ & 2 \\
T.J.\ Watt & 3 & $0.0348$ & $[0.0212,\;0.0410]$ & 4 \\
Myles Garrett & 4 & $0.0342$ & $[0.0211,\;0.0406]$ & 5 \\
Trey Hendrickson & 5 & $0.0342$ & $[0.0225,\;0.0367]$ & 39 \\
\midrule
\multicolumn{5}{l}{\textit{Offensive linemen ($n=177$)}} \\
Dion Dawkins & 1 & $0.0130$ & $[0.0109,\;0.0178]$ & 40 \\
Halapoulivaati Vaitai & 2 & $0.0131$ & $[0.0114,\;0.0181]$ & 73 \\
Corey Linsley & 3 & $0.0136$ & $[0.0112,\;0.0192]$ & 117 \\
Kolton Miller & 4 & $0.0136$ & $[0.0113,\;0.0179]$ & 9 \\
Tristan Wirfs & 5 & $0.0137$ & $[0.0116,\;0.0192]$ & 114 \\
\bottomrule
\end{tabular}
\end{table}

For example, Myles Garrett and Joe Tryon-Shoyinka both rank among the top ten for initial matchup strength, but their full-play disruption ranks are fourth and 53rd, respectively.
Their early pass rush win rates in the final-model sample are similar (31.1\% and 31.5\%), yet Garrett recorded pressures more often and accumulated nearly three times as much sack credit per matchup.

\FloatBarrier

\subsection{All-Pro Alignment and Sensitivity}\label{sec:results_comparisons}
Among players with at least 200 eligible matchups, the fitted models place more 2021 All-Pro recipients in their top-ten rankings than either the raw or smoothed baselines, for both rushers and offensive linemen (Table~\ref{tab:all_pro_alignment}).
For full-play disruption, either-team AUC is also higher than both baselines in each role, while first-team AUC is similar for rushers and lower for linemen.
For early pass rush outcomes, AUC is higher for rushers but lower for offensive linemen.

\begin{table}[htbp]
\centering
\small
\renewcommand{\arraystretch}{0.95}
\caption{All-Pro alignment of fitted, raw, and smoothed ratings among players with at least 200 eligible matchups. Each group uses the same players across methods. Highest values across methods are bold, including ties.}
\label{tab:all_pro_alignment}
\begin{tabular}{lrrrr}
\toprule
& \multicolumn{2}{c}{AUC} & \multicolumn{2}{c}{All-Pro in top ten} \\
\cmidrule(lr){2-3}\cmidrule(lr){4-5}
Method & First team & Either team & First team & Either team \\
\midrule
\multicolumn{5}{l}{\textit{Early pass rush: rushers ($n=139$)}} \\
Fitted & \textbf{0.807} & \textbf{0.815} & \textbf{4} & \textbf{6} \\
Raw & 0.785 & 0.770 & 3 & 4 \\
Smoothed & 0.785 & 0.770 & 3 & 4 \\
\midrule
\multicolumn{5}{l}{\textit{Early pass rush: offensive linemen ($n=177$)}} \\
Fitted & 0.614 & 0.657 & \textbf{2} & \textbf{3} \\
Raw & 0.627 & \textbf{0.666} & 1 & 1 \\
Smoothed & \textbf{0.628} & \textbf{0.666} & 1 & 1 \\
\midrule
\multicolumn{5}{l}{\textit{Full-play disruption: rushers ($n=138$)}} \\
Fitted & \textbf{0.911} & \textbf{0.915} & \textbf{4} & \textbf{5} \\
Raw & 0.910 & 0.876 & 3 & 4 \\
Smoothed & 0.910 & 0.876 & 3 & 4 \\
\midrule
\multicolumn{5}{l}{\textit{Full-play disruption: offensive linemen ($n=177$)}} \\
Fitted & 0.723 & \textbf{0.823} & \textbf{2} & \textbf{4} \\
Raw & 0.762 & 0.786 & 1 & 2 \\
Smoothed & \textbf{0.763} & 0.786 & 1 & 2 \\
\bottomrule
\end{tabular}
\par\smallskip
\begin{minipage}{\linewidth}
\footnotesize
Either team includes first- and second-team selections. Each rusher cohort contains five first-team and nine either-team recipients; each offensive-line cohort contains five and ten, respectively.
\end{minipage}
\end{table}

\FloatBarrier
At the 200-matchup cutoff, the proportional-odds model retains eight of the multinomial model's ten leading rushers, but only three of its ten leading offensive linemen.
Across all players at that cutoff, the absolute rank change averages about 18 places for rushers and 27 for linemen (Appendix~\ref{app:ordinal_model}).

The fitted models' top-ten advantage over both baselines also holds with a 100-matchup requirement when selections to either team are counted (Appendix~\ref{app:all_pro_comparisons}).
At 400 matchups, the rusher cohorts shrink to 13 players for early ratings and 11 for full-play disruption, with all three methods placing four All-Pro rushers in the top ten within each cohort.

\Needspace{\textheight}
\subsection{Ratings Through the Season}\label{sec:results_weekly}
\begin{figure}[!b]
\centering
\includegraphics[width=\linewidth]{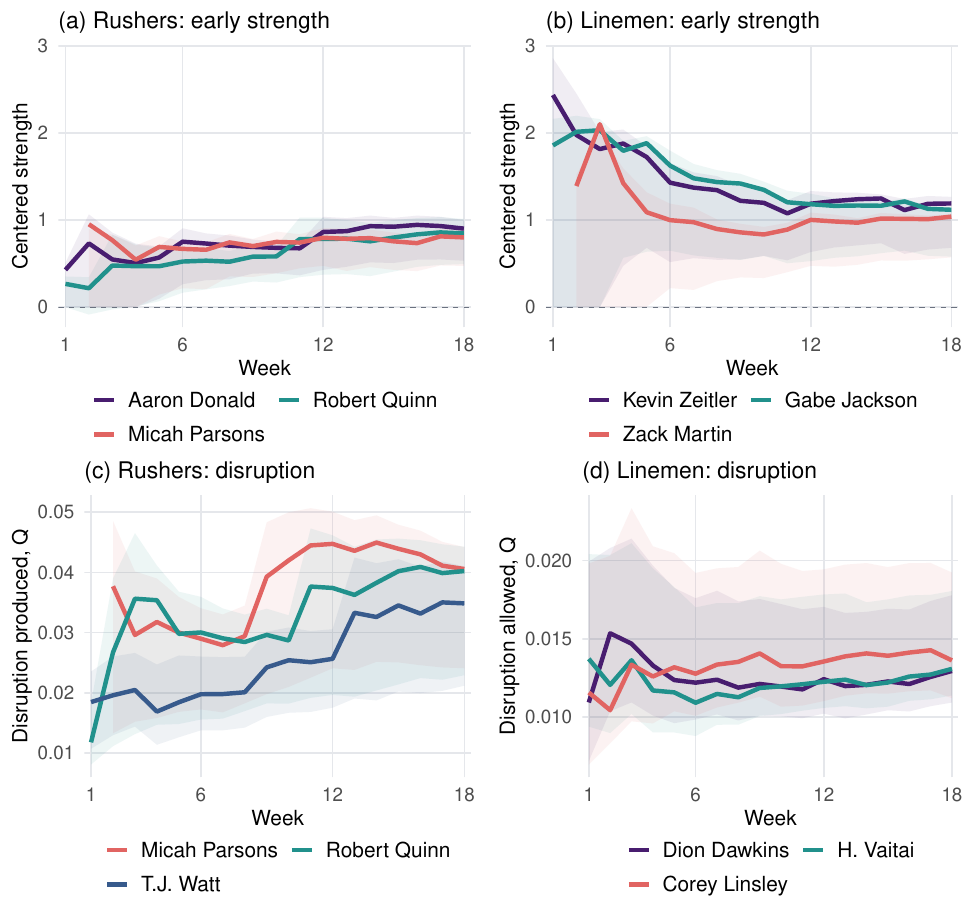}
\caption{Retrospective weekly trajectories of each model's three highest-rated players with at least 200 eligible full-season matchups. Panels (a,b) show centered early pass rush strength for rushers and offensive linemen; (c,d) show their expected full-play disruption, $Q$. Players are selected separately by model and role based on their Week 18 ratings. Shading shows players' 95\% pointwise game-bootstrap percentile intervals from all 1,000 draws. Higher strength is favorable in both roles, while higher $Q$ is favorable for rushers and unfavorable for linemen.}
\label{fig:weekly_leader_overlays}
\end{figure}

Weekly ratings trace how our assessments of players develop over the season, with bootstrap bands showing the uncertainty along each trajectory (Figure~\ref{fig:weekly_leader_overlays}).
For example, Aaron Donald's early pass rush rating rises from Week 5 to season end while Myles Garrett's declines, reversing their point ordering by Week 12.
Early pass rush and disruption ratings need not move together: between Weeks 8 and 10, Micah Parsons' disruption rating increases from 0.029 to 0.042, while his early pass rush rating is near 0.75 at both weeks.

\FloatBarrier

\section{Discussion}\label{sec:discussion}
\subsection{Implications for Player Evaluation}\label{sec:discussion_implications}
By estimating blocker and rusher strength jointly, our framework controls for opponent strength and protection help when evaluating individual players.
These adjustments put a player's observed success in the context of his assignments, giving credit to linemen who sustain protection against consistently stronger rushers or with less help.
The gains in held-out prediction demonstrate the framework's predictive value beyond historical outcome rates, while the greater concentration of All-Pro recipients among the fitted top-ten players adds descriptive support for these evaluations.

Within that framework, early pass rush strength and full-play disruption answer different football questions; early ratings concern performance in the initial blocking assignment, while full-play ratings emphasize the pressures and sacks produced or allowed and their estimated value.
Which rating we emphasize therefore shapes what we reward in player evaluation, and this choice is especially important for offensive linemen, for whom the two models identify substantially different leaders.

Uncertainty also shapes how these comparisons should be used.
The closely spaced early strength estimates for offensive linemen make their relative ordering harder to resolve, even when individual rating intervals are similar in width to those for rushers.
Their full-play disruption rankings are also more sensitive to model specification, so sampling variation and modeling choices both affect confidence in an exact rank.
Player comparisons should therefore draw on the estimated ratings and their intervals, alongside their sensitivity to outcome modeling.
Weekly paths add the timing of revisions to those assessments, with uncertainty shown at each stage.

\subsection{Limitations}\label{sec:discussion_limitations}
As noted in Section~\ref{sec:outcomes}, quick pass releases and short initial matchups reduce the time available to observe an early pass rush win, so a recorded loss does not necessarily imply protection sustained for the full 2.5 seconds.
Additionally, full-play outcomes present a separate attribution problem: we associate pressures and sacks with the initial blocking group even when later pickups, quarterback behavior, or other players contribute to the result.

These records also shape which assignments our findings represent.
If tracking or annotations are missing more often in particular situations, the retained matchups and complete EPA plays can underrepresent those situations, potentially affecting both player ratings and disruption weights.
With data from 2021 alone, the stability of these findings across seasons also remains untested.

Even within the retained matchups, the models simplify how blockers share responsibility.
For example, a primary blocker and a teammate providing a brief assist receive equal weight in an assignment's strength, and the help indicator does not distinguish a chip from sustained double-team protection.
Other differences in protection scheme, route timing, or quarterback movement remain unmeasured and may enter the player effects, so the ratings can incorporate scheme and usage alongside individual performance.
Although the bootstrap intervals capture sampling and refitting variation, they do not quantify uncertainty from outcome recording, sample selection, or model assumptions.

On the disruption scale, the EPA weights describe associations after adjustment for recorded play context, without isolating points caused by a player's actions.
The resulting ratings summarize expected disruption under a common mix of blocking assignments drawn from the observed season, so their interpretation is specific to that reference.
That reference remains fixed in the weekly paths, alongside season-level tuning and EPA weights.
These choices incorporate information unavailable at some earlier weeks, making the paths retrospective assessments.

\subsection{Future Directions}\label{sec:discussion_future}
Validation across additional seasons would test how well the ratings predict subsequent performance and how stable the EPA-based outcome weights remain as personnel and protection schemes change.
Fitting on earlier seasons and predicting later games would be especially informative for players who change teams or responsibilities.

Richer descriptions of those responsibilities could also improve how the models represent shared protection.
Using block durations and handoff timing, together with information on each blocker's assigned role, would allow a primary block to be distinguished from a brief assist and protection to be followed beyond the initial matchup.
Combined with longer player histories, these details could support models that let strength vary over time, helping distinguish changes in performance from revisions driven by accumulating matchups.

To evaluate such ratings prospectively, each weekly fit would need to use only information available at the time.
Penalty selection, EPA calibration, and the player pool would therefore be determined from past observations, with a comparison reference also chosen from earlier data.
Keeping that reference fixed could support comparisons across weeks, while predictions for subsequent games would provide a test against the historical-rate baselines.

\section*{Reproducibility}
Code to reproduce our analysis is available at \url{https://github.com/WhartonSABI/nfl-elo}.
The play-by-play data are publicly available through nflverse \citep{nflverse2026}; reconstructing the blocking matchups requires access to the Hudl records described in Section~\ref{sec:data}.

\Needspace{0.27\textheight}
\section*{Acknowledgements}
We gratefully acknowledge the Wharton Sports Analytics and Business Initiative for supporting this work.
We thank Hudl for access to the 2021 dataset and Dr. Paul Sabin for early conversations that helped shape the project.

OpenAI's ChatGPT/Codex assisted with programming and figure preparation.
The authors reviewed the outputs and take responsibility for the final work.


\clearpage
\appendix
\begin{center}
  {\Large\bfseries Appendix}
\end{center}

\section{EPA Regression}\label{app:epa_model}
The compact notation in Equation~\ref{eq:epa_model} represents the linear regression
\begin{equation}\label{eq:epa_model_full}
\begin{aligned}
\mathrm{EPA}_\ell
={}&\beta_0+\beta_{\mathrm{win}}W_\ell+\beta_{\mathrm{pressure}}P_\ell+\beta_{\mathrm{sack}}S_\ell \\
&+\gamma_B B_\ell+\gamma_E E_\ell+\mathbf{x}_\ell^\top\boldsymbol{\theta}
+u_{q(\ell)}+v_{d(\ell)}+\varepsilon_\ell.
\end{aligned}
\end{equation}
We fit this regression by ordinary least squares on the training plays described in Section~\ref{sec:valuation}.
The terms $W_\ell$ and $P_\ell$ count defenders in the win and pressure categories of the final pass rush outcome, while $S_\ell$ sums sack credits.
Pressure outcomes and sack credits include recorded events from defenders without an observed initial matchup.

The count $B_\ell$ is the number of rushers with an observed initial blocking assignment.
To account for shared sacks, $E_\ell$ is the number of sack recipients among those rushers minus their combined sack credit.
For example, two such rushers with half sacks contribute $S_\ell=1$ and $E_\ell=1$.

The context vector $\mathbf{x}_\ell$ includes indicators for down, the second half, and overtime, together with the logarithm of one plus yards to go, field position, seconds remaining in the half, and score differential.
The terms $u_{q(\ell)}$ and $v_{d(\ell)}$ are quarterback and defensive-team effects.

\section{Cross-Validation Tuning}\label{app:cv_tuning}
\begin{figure}[htbp]
\centering
\includegraphics[width=0.92\linewidth]{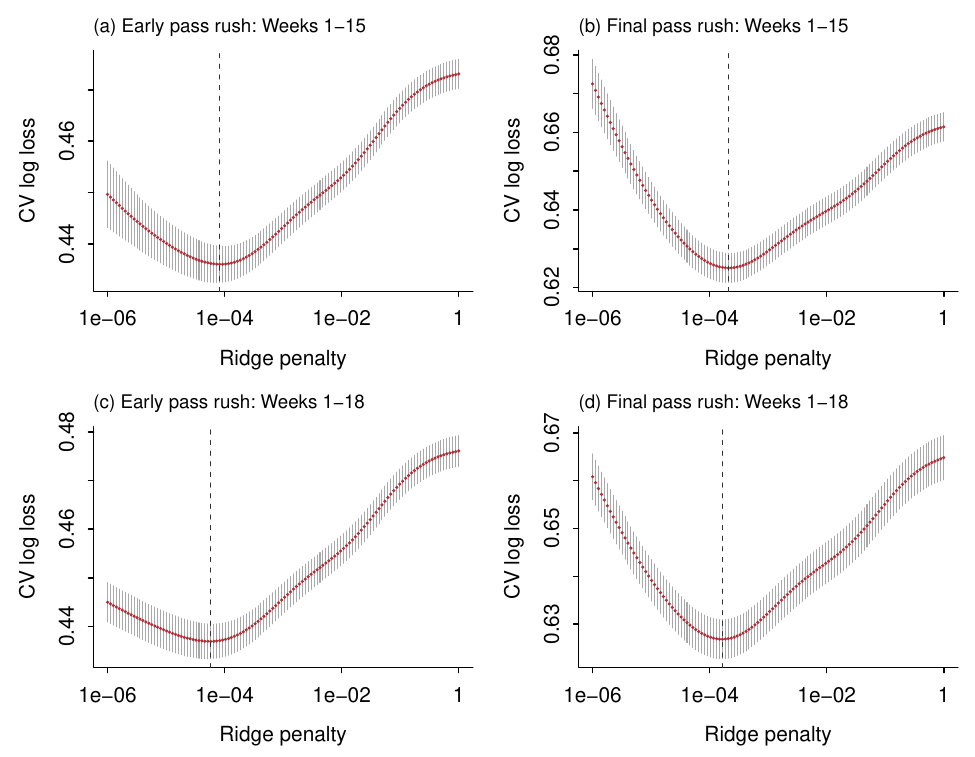}
\caption{Five-fold game-grouped cross-validation curves from \texttt{cv.glmnet} over 120 logarithmically spaced ridge penalties from $10^{-6}$ to 1. Panels (a,b) show the early and final pass rush models fitted on Weeks 1--15 for predictive evaluation; panels (c,d) show the corresponding full-season fits on Weeks 1--18 for player ratings. Points give mean cross-validation log loss, bars extend one cross-validation standard error above and below each mean, and dashed vertical lines mark the minimum-loss penalty used in each fit. These bars summarize variation across validation folds and are distinct from the game-bootstrap intervals for player ratings.}
\label{fig:ridge_cv_curves}
\end{figure}

Figure~\ref{fig:ridge_cv_curves} displays cross-validation performance across the ridge penalty grid used in Section~\ref{sec:estimation_tuning}.
Training-period tuning supports predictive evaluation, while full-season tuning determines the penalties used for season ratings.

\FloatBarrier


\Needspace{8\baselineskip}
\section{Prediction Baselines}\label{app:predictive_baselines}
For either pass rush model, let $g_c$ denote the frequency of outcome $c$ among eligible training matchups; the league-average baseline assigns these probabilities to every test matchup.

To build the historical matchup baseline, let $n_{a,c}$ denote player $a$'s training count for outcome $c$, with $n_a=\sum_c n_{a,c}$.
Each rusher contributes one count per matchup, while each blocker in a shared assignment of size $m_k$ contributes $1/m_k$.
We smooth these counts to obtain the outcome profile
\begin{equation}\label{eq:baseline_smoothing}
h_{a,c}=\frac{n_{a,c}+\tau g_c}{n_a+\tau},
\end{equation}
where $\tau$ is chosen separately for each model by five-fold game-grouped cross-validation within Weeks 1--15.
We consider $\tau\in\{0,1,2,5,10,25,50,100,200,400,1000,\infty\}$, with $0$ giving unsmoothed profiles and $\infty$ giving the league profile.
Each fold rebuilds the league and player histories from its training games, and we select the strength with the lowest average validation log loss.
A player without a training history also receives $g_c$.

We then form the prediction for matchup $k$ by taking the geometric mean of the blockers' profiles, combining it with the rusher's profile, and normalizing across outcomes:
\begin{equation}\label{eq:baseline_pooling}
\begin{aligned}
h_{J_k,c}&=\left(\prod_{j\in J_k}h_{j,c}\right)^{1/m_k},\\
\widehat{\pi}^{\mathrm{hist}}_{k,c}
&=\frac{\sqrt{h_{i(k),c}h_{J_k,c}}}
{\sum_d\sqrt{h_{i(k),d}h_{J_k,d}}}.
\end{aligned}
\end{equation}

We apply a numerical probability floor of $10^{-15}$ to blocker profiles before geometric pooling and to log-loss evaluation, including the unsmoothed endpoint.

The training sample selects $\tau=2$ for early pass rush outcomes and $\tau=10$ for final outcomes.
The primary comparisons in Table~\ref{tab:predictive_performance} retain these strengths throughout the game bootstrap, rebuilding the histories in each draw.
Alongside them, we report a retuned comparison that reselects smoothing within each draw's training games.
Appendix~\ref{app:baseline_sensitivity} gives the comparisons across the full fixed grid using the same draws.

For descriptive season ranking comparisons, we rebuild the player profiles from all eligible season matchups, using $n_{a,c}/n_a$ for raw rates and Equation~\ref{eq:baseline_smoothing} for smoothed rates with the original Weeks 1--15 cross-validation choices: 2 for early outcomes and 10 for final outcomes.
A blocker's initial protection rate is one minus the early pass rush win frequency in his profile.
Applying the EPA weights to each final-outcome profile gives the disruption rating, with lower values indicating better protection.
For these comparisons, exposure thresholds count each player appearance once, including shared assignments, rather than using the allocated count $n_a$.
Tied scores receive average ranks, while ties at the top-ten boundary divide the remaining places equally.

\Needspace{0.70\textheight}
\section{Ordinal Sensitivity}\label{app:ordinal_model}
Using the final pass rush model's eligible matchups, we fit a proportional-odds model with ordered outcomes $\mathrm{loss}<\mathrm{win}<\mathrm{pressure}<\mathrm{sack}$:
\begin{equation}\label{eq:ordinal_model}
\begin{aligned}
\operatorname{logit}\Pr(C_k\le c)&=\kappa_c-\eta_k^{\mathrm{ord}},\\
\eta_k^{\mathrm{ord}}&=r_{i(k)}^{\mathrm{ord}}-\frac{1}{m_k}\sum_{j\in J_k}b_j^{\mathrm{ord}}
+\delta^{\mathrm{ord}}D_k+\xi^{\mathrm{ord}}U_k,
\end{aligned}
\end{equation}
where $c$ indexes the first three categories, $\kappa_{\mathrm{loss}}<\kappa_{\mathrm{win}}<\kappa_{\mathrm{pressure}}$, and $U_k$ indicates an unavailable help annotation.
The thresholds absorb the intercept, while the player and help effects are shared across thresholds.
After selecting the ridge penalty by five-fold game-grouped cross-validation on the full season, we refit the model and calculate ratings as in Section~\ref{sec:ratings}, retaining the multinomial model's EPA weights and reference assignments.
For the ranking comparisons in Table~\ref{tab:ordinal_rank_sensitivity}, we restrict rushers to defensive linemen and linebackers and blockers to offensive linemen, applying the exposure thresholds from Section~\ref{sec:player_comparisons}.

\begin{table}[htbp]
\centering
\small
\caption{Sensitivity of full-play disruption rankings to the proportional-odds specification. Rankings use the same eligible players within each role and exposure group. A zero cutoff retains players with positive observed exposure.}
\label{tab:ordinal_rank_sensitivity}
\begin{tabular}{rrrrr}
\toprule
\shortstack{Minimum\\matchups} & Players & Spearman & \shortstack{Mean absolute\\rank change} & \shortstack{Top-ten\\overlap} \\
\midrule
\multicolumn{5}{l}{\textit{Rushers (defensive linemen and linebackers)}} \\
0 & 603 & 0.871 & 57.4 & 7 \\
100 & 251 & 0.849 & 28.4 & 7 \\
200 & 138 & 0.796 & 18.2 & 8 \\
400 & 11 & 0.918 & 0.9 & 9 \\
\midrule
\multicolumn{5}{l}{\textit{Offensive linemen}} \\
0 & 339 & 0.817 & 44.2 & 3 \\
100 & 222 & 0.776 & 32.4 & 3 \\
200 & 177 & 0.753 & 27.0 & 3 \\
400 & 88 & 0.731 & 14.0 & 3 \\
\bottomrule
\end{tabular}
\end{table}

\FloatBarrier
\section{All-Pro Comparisons}\label{app:all_pro_comparisons}
Tables~\ref{tab:all_pro_exposure} and~\ref{tab:all_pro_exposure_first} extend the baseline comparisons in Table~\ref{tab:all_pro_alignment} across exposure requirements, using either-team and first-team All-Pro selections, respectively.


\begin{table}[H]
\centering
\small
\caption{All-Pro alignment of fitted, raw, and smoothed ratings across minimum matchup requirements, using selections to either All-Pro team. All cohorts require positive observed exposure, including those with a zero cutoff. Highest values across methods are bold, including ties.}
\label{tab:all_pro_exposure}
\begin{tabular}{rrrrrrrr}
\toprule
& & \multicolumn{3}{c}{AUC} & \multicolumn{3}{c}{All-Pro in top ten} \\
\cmidrule(lr){3-5}\cmidrule(lr){6-8}
\shortstack{Minimum\\matchups} & Players & Fitted & Raw & Smoothed & Fitted & Raw & Smoothed \\
\midrule
\multicolumn{8}{l}{\textit{Early pass rush: rushers}} \\
0 & 604 & \textbf{0.756} & 0.726 & 0.732 & \textbf{5} & 0 & 0 \\
100 & 253 & \textbf{0.825} & 0.782 & 0.782 & \textbf{5} & 4 & 4 \\
200 & 139 & \textbf{0.815} & 0.770 & 0.770 & \textbf{6} & 4 & 4 \\
400 & 13 & \textbf{0.675} & 0.650 & 0.650 & \textbf{4} & \textbf{4} & \textbf{4} \\
\midrule
\multicolumn{8}{l}{\textit{Early pass rush: offensive linemen}} \\
0 & 339 & \textbf{0.736} & 0.666 & 0.690 & \textbf{3} & 0 & 0 \\
100 & 224 & 0.673 & \textbf{0.677} & \textbf{0.677} & \textbf{3} & 1 & 1 \\
200 & 177 & 0.657 & \textbf{0.666} & \textbf{0.666} & \textbf{3} & 1 & 1 \\
400 & 91 & 0.606 & \textbf{0.639} & \textbf{0.639} & \textbf{1} & \textbf{1} & \textbf{1} \\
\midrule
\multicolumn{8}{l}{\textit{Full-play disruption: rushers}} \\
0 & 603 & \textbf{0.792} & 0.768 & 0.744 & \textbf{5} & 0 & 3 \\
100 & 251 & \textbf{0.931} & 0.891 & 0.893 & \textbf{5} & 4 & 4 \\
200 & 138 & \textbf{0.915} & 0.876 & 0.876 & \textbf{5} & 4 & 4 \\
400 & 11 & \textbf{0.750} & \textbf{0.750} & \textbf{0.750} & \textbf{4} & \textbf{4} & \textbf{4} \\
\midrule
\multicolumn{8}{l}{\textit{Full-play disruption: offensive linemen}} \\
0 & 339 & \textbf{0.877} & 0.695 & 0.780 & \textbf{3} & 0 & 1 \\
100 & 222 & \textbf{0.837} & 0.790 & 0.792 & \textbf{3} & 1 & 1 \\
200 & 177 & \textbf{0.823} & 0.786 & 0.786 & \textbf{4} & 2 & 2 \\
400 & 88 & \textbf{0.831} & 0.748 & 0.748 & \textbf{4} & 1 & 1 \\
\bottomrule
\end{tabular}
\end{table}

\begin{table}[H]
\centering
\small
\caption{All-Pro alignment of fitted, raw, and smoothed ratings across minimum matchup requirements, using first-team All-Pro selections. All cohorts require positive observed exposure, including those with a zero cutoff. Highest values across methods are bold, including ties.}
\label{tab:all_pro_exposure_first}
\begin{tabular}{rrrrrrrr}
\toprule
& & \multicolumn{3}{c}{AUC} & \multicolumn{3}{c}{All-Pro in top ten} \\
\cmidrule(lr){3-5}\cmidrule(lr){6-8}
\shortstack{Minimum\\matchups} & Players & Fitted & Raw & Smoothed & Fitted & Raw & Smoothed \\
\midrule
\multicolumn{8}{l}{\textit{Early pass rush: rushers}} \\
0 & 604 & 0.759 & 0.759 & \textbf{0.765} & \textbf{4} & 0 & 0 \\
100 & 253 & \textbf{0.818} & 0.795 & 0.795 & \textbf{4} & 3 & 3 \\
200 & 139 & \textbf{0.807} & 0.785 & 0.785 & \textbf{4} & 3 & 3 \\
400 & 13 & \textbf{0.667} & 0.633 & 0.633 & \textbf{2} & \textbf{2} & \textbf{2} \\
\midrule
\multicolumn{8}{l}{\textit{Early pass rush: offensive linemen}} \\
0 & 339 & \textbf{0.681} & 0.635 & 0.654 & \textbf{2} & 0 & 0 \\
100 & 224 & 0.628 & \textbf{0.637} & \textbf{0.637} & \textbf{2} & 1 & 1 \\
200 & 177 & 0.614 & 0.627 & \textbf{0.628} & \textbf{2} & 1 & 1 \\
400 & 91 & \textbf{0.659} & 0.625 & 0.625 & \textbf{1} & \textbf{1} & \textbf{1} \\
\midrule
\multicolumn{8}{l}{\textit{Full-play disruption: rushers}} \\
0 & 603 & \textbf{0.837} & 0.797 & 0.763 & \textbf{4} & 0 & 2 \\
100 & 251 & \textbf{0.927} & 0.922 & 0.924 & \textbf{4} & 3 & 3 \\
200 & 138 & \textbf{0.911} & 0.910 & 0.910 & \textbf{4} & 3 & 3 \\
400 & 11 & 0.611 & \textbf{0.778} & \textbf{0.778} & \textbf{2} & \textbf{2} & \textbf{2} \\
\midrule
\multicolumn{8}{l}{\textit{Full-play disruption: offensive linemen}} \\
0 & 339 & \textbf{0.796} & 0.674 & 0.756 & \textbf{1} & 0 & 0 \\
100 & 222 & 0.740 & 0.765 & \textbf{0.767} & \textbf{1} & 0 & 0 \\
200 & 177 & 0.723 & 0.762 & \textbf{0.763} & \textbf{2} & 1 & 1 \\
400 & 88 & \textbf{0.737} & 0.686 & 0.686 & \textbf{2} & 0 & 0 \\
\bottomrule
\end{tabular}
\end{table}

\FloatBarrier
\Needspace{0.88\textheight}
\section{Weekly Rating Summaries}\label{app:weekly_summaries}
\begin{table}[H]
\centering
\small
\caption{Spearman correlations between weekly ratings and the corresponding Week 18 ratings. The fixed season cohorts contain players with at least 200 eligible matchups; at each week, we compare those who have already appeared in an eligible matchup. Parentheses give the number of players compared.}
\label{tab:weekly_rank_alignment}
\begin{tabular}{llrrrr}
\toprule
Model & Players & Season $n$ & Week 5 & Week 10 & Week 15 \\
\midrule
Early pass rush & Rushers & 139 & 0.762 (138) & 0.891 (139) & 0.971 (139) \\
Early pass rush & Offensive linemen & 177 & 0.611 (168) & 0.848 (176) & 0.958 (177) \\
Full-play disruption & Rushers & 138 & 0.597 (137) & 0.823 (138) & 0.937 (138) \\
Full-play disruption & Offensive linemen & 177 & 0.591 (168) & 0.803 (176) & 0.933 (177) \\
\bottomrule
\end{tabular}
\end{table}

\begin{table}[H]
\centering
\small
\caption{Cumulative eligible matchup counts for the five full-season leaders in each model and role.}
\label{tab:weekly_leader_counts}
\begin{tabular}{lrrrr}
\toprule
Player & Week 4 & Week 8 & Week 12 & Week 18 \\
\midrule
\multicolumn{5}{l}{\textit{Rushers: early pass rush strength}} \\
Aaron Donald & 99 & 206 & 296 & 489 \\
Robert Quinn & 79 & 139 & 200 & 298 \\
Micah Parsons & 57 & 83 & 161 & 210 \\
T.J. Watt & 68 & 141 & 181 & 315 \\
Myles Garrett & 92 & 184 & 269 & 418 \\
\multicolumn{5}{l}{\textit{Rushers: full-play disruption}} \\
Micah Parsons & 57 & 83 & 161 & 210 \\
Robert Quinn & 79 & 139 & 200 & 298 \\
T.J. Watt & 68 & 120 & 160 & 294 \\
Myles Garrett & 92 & 162 & 247 & 396 \\
Trey Hendrickson & 110 & 154 & 204 & 303 \\
\multicolumn{5}{l}{\textit{Offensive linemen: early pass rush strength}} \\
Kevin Zeitler & 114 & 216 & 356 & 581 \\
Gabe Jackson & 94 & 189 & 285 & 431 \\
Zack Martin & 69 & 179 & 330 & 489 \\
Lane Johnson & 79 & 112 & 190 & 285 \\
Trent Williams & 112 & 156 & 250 & 380 \\
\multicolumn{5}{l}{\textit{Offensive linemen: full-play disruption}} \\
Dion Dawkins & 136 & 231 & 318 & 461 \\
Halapoulivaati Vaitai & 119 & 236 & 261 & 407 \\
Corey Linsley & 147 & 253 & 392 & 513 \\
Kolton Miller & 166 & 258 & 393 & 551 \\
Tristan Wirfs & 33 & 172 & 260 & 472 \\
\bottomrule
\end{tabular}
\end{table}

\FloatBarrier

\begin{figure}[H]
\centering
\includegraphics[width=\linewidth]{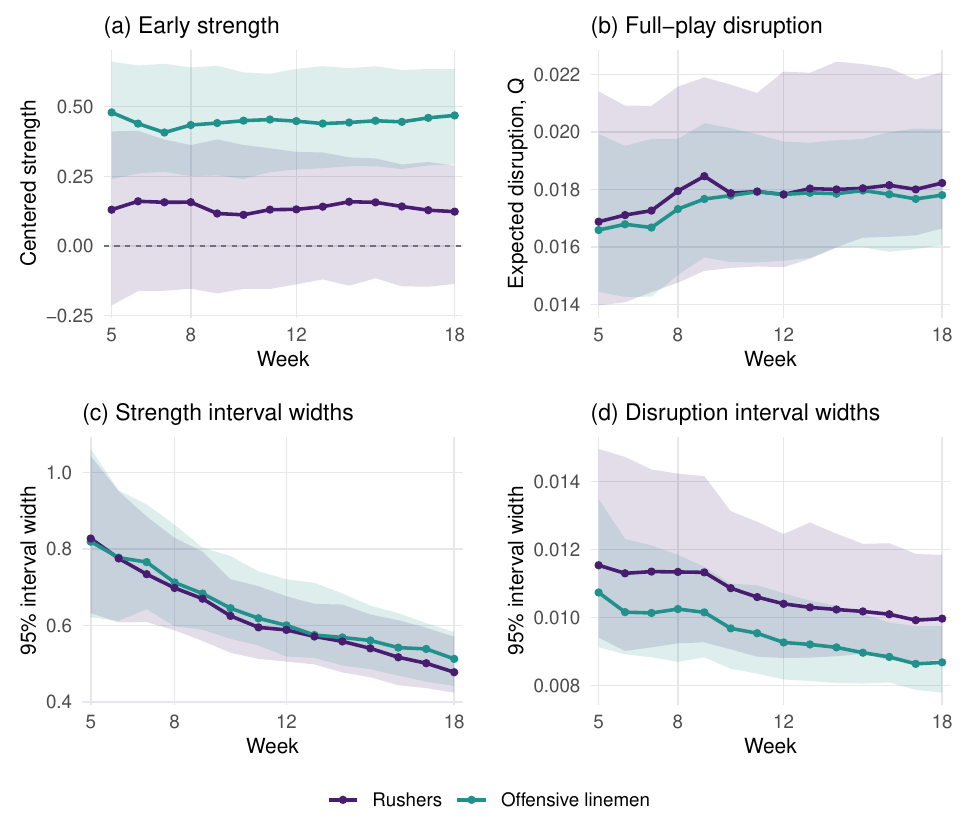}
\caption{Retrospective weekly distributions of fitted ratings and individual interval widths. Lines show cohort medians and shading shows the IQR across players. Panels (a,b) summarize centered early pass rush strength and full-play disruption, $Q$; (c,d) summarize widths of players' 95\% pointwise game-bootstrap percentile intervals. Within each role, both models follow the same players: those with at least 200 eligible matchups in each model and at least one in each by Week 5.}
\label{fig:weekly_cohort_summary}
\end{figure}

\FloatBarrier

\FloatBarrier
\Needspace{0.58\textheight}
\section{Baseline Smoothing Sensitivity}\label{app:baseline_sensitivity}
Table~\ref{tab:baseline_sensitivity} compares held-out prediction across the smoothing grid in Appendix~\ref{app:predictive_baselines}.
The fixed-strength rows retain the indicated value in every bootstrap draw; for the retuned comparison, we select the strength with the lowest cross-validation loss in that draw's training games.
All comparisons use the same fitted-model losses and sampled test games.

\begin{table}[H]
\centering
\small
\caption{Percentage reductions in held-out log loss relative to historical matchup baselines across smoothing strengths, with 95\% game-bootstrap percentile intervals. Bold reductions mark the primary fixed choices of 2 for early outcomes and 10 for final outcomes. For the retuned comparison, we select the strength with the lowest cross-validation loss in each draw's training games.}
\label{tab:baseline_sensitivity}
\begin{tabular}{lrrrr}
\toprule
 & \multicolumn{2}{c}{Early pass rush} & \multicolumn{2}{c}{Final pass rush} \\
Strength & Reduction (\%) & 95\% interval & Reduction (\%) & 95\% interval \\
\midrule
0 & 33.0 & $[27.4,\;48.5]$ & 39.0 & $[41.9,\;53.6]$ \\
1 & 4.6 & $[3.5,\;5.6]$ & 4.4 & $[4.7,\;6.5]$ \\
2 & \textbf{4.5} & $[3.4,\;5.2]$ & 4.0 & $[4.1,\;5.7]$ \\
5 & 4.6 & $[3.4,\;5.0]$ & 3.7 & $[3.5,\;4.8]$ \\
10 & 4.8 & $[3.5,\;5.0]$ & \textbf{3.6} & $[3.2,\;4.4]$ \\
25 & 5.2 & $[3.8,\;5.3]$ & 3.7 & $[3.1,\;4.2]$ \\
50 & 5.7 & $[4.2,\;5.7]$ & 3.9 & $[3.1,\;4.2]$ \\
100 & 6.2 & $[4.7,\;6.2]$ & 4.3 & $[3.3,\;4.4]$ \\
200 & 6.8 & $[5.2,\;6.8]$ & 4.6 & $[3.5,\;4.7]$ \\
400 & 7.4 & $[5.7,\;7.4]$ & 5.1 & $[3.8,\;5.1]$ \\
1000 & 8.1 & $[6.3,\;8.1]$ & 5.5 & $[4.2,\;5.6]$ \\
$\infty$ & 8.8 & $[7.0,\;9.0]$ & 6.1 & $[4.7,\;6.3]$ \\
\midrule
Retuned & 4.5 & $[3.4,\;5.0]$ & 3.6 & $[3.1,\;4.2]$ \\
\bottomrule
\end{tabular}
\end{table}

\FloatBarrier



\begin{thebibliography}{9}
\bibitem[Alamar and Weinstein-Gould(2008)]{alamar2008}
Alamar, B. C., and Weinstein-Gould, J. (2008).
Isolating the effect of individual linemen on the passing game in the National Football League.
\textit{Journal of Quantitative Analysis in Sports}, 4(2).
\url{https://doi.org/10.2202/1559-0410.1113}.

\bibitem[Associated Press(2022)]{apallpro2021}
Associated Press (2022).
2021 NFL All-Pro Team.
\textit{NFL.com}.
\url{https://www.nfl.com/news/2021-all-pro-team-jonathan-taylor-t-j-watt-cooper-kupp-highlight-roster}.

\bibitem[Bradley and Terry(1952)]{bradley1952}
Bradley, R. A., and Terry, M. E. (1952).
Rank analysis of incomplete block designs: I. The method of paired comparisons.
\textit{Biometrika}, 39(3--4), 324--345.
\url{https://doi.org/10.1093/biomet/39.3-4.324}.

\bibitem[Burke(2018)]{burke2018}
Burke, B. (2018).
We created better pass-rusher and pass-blocker stats: How they work.
\textit{ESPN Analytics}, October 5.
\url{https://www.espn.com/nfl/story/_/id/24892208/}.

\bibitem[Friedman et~al.(2010)Friedman, Hastie, and Tibshirani]{friedman2010}
Friedman, J., Hastie, T., and Tibshirani, R. (2010).
Regularization paths for generalized linear models via coordinate descent.
\textit{Journal of Statistical Software}, 33(1), 1--22.
\url{https://doi.org/10.18637/jss.v033.i01}.

\bibitem[National Football League(2021)]{pitbuf2021}
National Football League (2021).
\textit{Pittsburgh Steelers at Buffalo Bills: Gamebook, September 12, 2021}.
Official NFL gamebook, fourth-quarter play-by-play, p.~15.
\url{https://static.clubs.nfl.com/image/upload/steelers/a6awhlaxjc0yopdi0pec.pdf}.

\bibitem[National Football League(2022)]{bdb2023}
National Football League (2022).
NFL Big Data Bowl 2023.
\url{https://www.kaggle.com/competitions/nfl-big-data-bowl-2023/discussion/359079}.

\bibitem[nflverse(2026)]{nflverse2026}
nflverse (2026).
\textit{2021 NFL Play-by-Play Data}.
Frozen copy of the nflverse-data play-by-play release, retrieved September 16, 2026.
\url{https://github.com/nflverse/nflverse-data/releases/tag/pbp}.

\bibitem[Nguyen et~al.(2024)Nguyen, Yurko, and Matthews]{nguyen2023}
Nguyen, Q., Yurko, R., and Matthews, G. J. (2024).
Here comes the STRAIN: Analyzing defensive pass rush in American football with player tracking data.
\textit{The American Statistician}, 78(2), 199--208.
\url{https://doi.org/10.1080/00031305.2023.2242442}.

\end{thebibliography}
\end{document}